\documentclass[epj]{svjour}

\usepackage{amssymb}
\usepackage{graphicx}
\usepackage{dcolumn}
\usepackage{amsmath}

\begin{document}
\title{Pseudofermion ferromagnetism in the Kondo lattices: a mean-field
approach}
\author{V. Yu. Irkhin%
\thanks{Valentin.Irkhin@imp.uran.ru}%
}
\institute{Institute of Metal Physics, 620990 Ekaterinburg, Russia}

\abstract{
Ground state ferromagnetism of the Kondo lattices is investigated within slave fermion
approach by Coleman and Andrei within a mean-field approximation in the effective
hybridization model. Conditions
for formation of both saturated (half-metallic) and non-saturated
magnetic state are obtained for various lattices.
A description in terms of universal functions which depend only on bare electron
density of states (DOS) is
presented. A crucial role of
the energy dependence of the bare DOS (especially, of
DOS peaks) for the small-moment ferromagnetism formation is demonstrated.
}

\maketitle

\section{Introduction}

Experimental investigations of heavy-fermion and other anomalous $4f$- and $%
5f$-compounds, which are treated usually as Kondo lattices, demonstrate that
magnetic ordering is widely spread among such systems. There exist numerous
examples of systems where \textquotedblleft Kondo\textquotedblright\
anomalies in thermodynamic and transport properties coexist with
antiferromagnetic ordering and/or strong spin fluctuations. There are also
examples of Kondo ferromagnets: CeNiSb, CePdSb, CeSi$_{x}$, CeRh$_{3}$B$_{2}$%
, Sm$_{3}$Sb$_{4}$, NpAl$_{2}$ (see review and bibliography in \cite{IK,II}%
). The number of such materials gradually increases, including CePt \cite%
{CePt}, CeRu$_{2}$Ge$_{2}$ \cite{CeRuGe}, CeAgSb$_{2}$, \cite{CeAgSb}, CeRu$%
_{2}$M$_{2}$X (M = Al, Ga; X = B, C) \cite{CeRu2Ga2B1,CeRu2Ga2B}, CeIr$_{2}$B%
$_{2}$ \cite{CeIr2B2}, hydrogenated CeNiSn \cite{CeNiSn}. Among 2D-like
systems, we can mention CeRuPO, a ferromagnetic Kondo lattice where LSDA+U
calculations evidence a quasi-two-dimensional electronic band structure \cite%
{CeRuPO}. Recently ferromagnetic state with small magnetic moments was
investigated for the Kondo systems CeRuSi$_{2}$ \cite{CeRuSi2} and Ce$_{4}$Sb%
$_{1.5}$Ge$_{1.5}$ \cite{Ce4Sb3}.

Owing to competition with the Kondo effect, ferromagnetism of the Kondo
lattices has itself anomalous features (instability of magnetic state, small
values of magnetic moment and magnetic entropy, negative paramagnetic Curie
temperature etc.) \cite{IK,II}. Unusual nature of Kondo systems on the
border of magnetism was recently reviewed by Coleman \cite{Coleman1}.
Formation of the magnetic state, coexisting with the Kondo effect and
possessing a small magnetic moment, was treated in Ref.\cite{IK} within a
scaling approach. However, this method is insufficient to describe
quantitatively the ground state. An attempt was made to solve this problem
on the basis of mean-field approximation \cite{IK91} by using the
pseudofermion representation for localized $f$-spins \cite{Coleman}. In this
approach, the pseudofermions become delocalized and the system is described
by an effective $s-f$ hybridization model (Sect.2).

Unfortunately, only saturated ferromagnetic solutions were obtained in Ref.
\cite{IK91} since the bare density of states was taken in the rectangular
form.  In the present paper we consider the formation of the Kondo
ferromagnetism by detailed analysis and numerical solution of equations of
the mean-field approximation for an arbitrary $\rho (E)$. In Sect.3 we
investigate the conditions for existence of the saturated (half-metallic)
ferromagnetic solution, and in Sect.4 for non-saturated (small moment)
ferromagnetism. In Sect.5 we present results of numerical calculations for a
number of concrete two- and three-dimensional lattices.

\section{The model}

We start from the standard Hamiltonian of the $s-f$ exchange model
\begin{equation}
H=\sum_{\mathbf{k}\sigma
}t_{\mathbf{k}}c_{\mathbf{k}\sigma }^{\dagger }c_{\mathbf{k}\sigma
}-\sum_{\mathbf{q}}J_{\mathbf{q}}\mathbf{S}_{-\mathbf{q}}
\mathbf{S}_{\mathbf{q}}+H_{\text{int}}
 \label{eq:G.2}
\end{equation}
\[
H_{\text{int}}=-I\sum_{i\sigma \sigma ^{\prime
}}(\mathbf{S}_i\mbox{\boldmath$\sigma $}_{\sigma \sigma ^{\prime
}})c_{i\sigma }^{\dagger }c_{i\sigma ^{\prime }},
\]
where $c_{\mathbf{k}\sigma }^{\dagger }$  are conduction electron operators, $t_{\mathbf{k}}$ is the bare electron spectrum,
$\mathbf{S}_i$ are operators for localized spins,
{\boldmath$\sigma $} are the Pauli matrices, $I$ is the  $s-f$ exchange parameter.

A number of papers (see, e.g., \cite{Nolting}) treat ferromagnetism in the periodic $s-d(f)$ model starting from the equation-of motion method for the original Hamiltonian (\ref{eq:G.2}), so that the subsystems of the conduction electrons and $f$-spins remain uncoupled in Kondo's sense. However, for anomalous $f$-system where the Kondo compensation (singlet formation) should be considered in the zero order approximation, such an approach turns out to be inappropriate. Therefore we construct the mean-field approximation describing the Kondo ground state with the use of the Abrikosov pseudofermion representation for localized spins $S=1/2$
\begin{equation}
\mathbf{S}_i=\frac 12\sum_{\sigma \sigma ^{\prime }}f_{i\sigma
}^{\dagger }\mbox{\boldmath$\sigma $}_{\sigma \sigma ^{\prime
}}f_{i\sigma ^{\prime }}
 \label{eq:O.1}
\end{equation}
with the subsidiary condition
\[
f_{i\uparrow }^{\dagger }f_{i\uparrow }+f_{i\downarrow }^{\dagger
}f_{i\downarrow }=1.
\]
Making the saddle-point approximation for the path integral describing the
spin-fermion interacting system\cite{Coleman} one reduces the Hamiltonian of the $s-f$ exchange interaction to the effective hybridization term:
\begin{eqnarray}
-I\sum_{\sigma \sigma ^{\prime }}c_{i\sigma }^{\dagger }c_{i\sigma
^{\prime }}(\mbox{\boldmath$\sigma $}_{\sigma \sigma ^{\prime
}}\mathbf{S}_i-\frac 12\delta _{\sigma \sigma ^{\prime }}) \nonumber \\
\rightarrow f_i^{\dagger }V_ic_i+c_i^{\dagger }V_i^{\dagger
}f_i-\frac 1{2I}{\rm Sp}{} (V_iV_i^{\dagger }),
 \label{eq:O.2}
\end{eqnarray}%
where the vector notations are used
\[
f_i^{\dagger }=(f_{i\uparrow }^{\dagger },f_{i\downarrow
}^{\dagger }),\qquad c_i^{\dagger }=(c_{i\uparrow }^{\dagger
},c_{i\downarrow }^{\dagger }),
\]
$V$ is the effective hybridization matrix which is determined from a minimum
of the free energy. In the ferromagnetic state we have \cite{IK91}

\begin{eqnarray}
H-\mu \hat{n} &=&\sum_{\mathbf{k}\sigma }[(t_{\mathbf{k}}-\mu )c_{\mathbf{k}%
\sigma }^{\dagger }c_{\mathbf{k}\sigma }+W_{\sigma }f_{\mathbf{k}\sigma
}^{\dagger }f_{\mathbf{k}\sigma }  \nonumber \\
&&+V_{\sigma }(c_{\mathbf{k}\sigma }^{\dagger }f_{\mathbf{k}\sigma }+f_{%
\mathbf{k}\sigma }^{\dagger }c_{\mathbf{k}\sigma })]-\sum_{\sigma }V_{\sigma
}^{2}/I \label{eq:O.4}
\end{eqnarray}%
with
\[
W_{\sigma }=W-\sigma J_{0}\bar{S},
\]%
$W$ being the energy of pseudofermion \textquotedblleft $f$%
-level\textquotedblright , $J_{0}=J(\mathbf{{q}=0)}$ the maximum Fourier transfor of the Heisenberg interaction.
Thus we reduce the periodic Kondo ($s-f$ exchange) model to an effective
hybridization model with the spin-dependent parameters $W_{\sigma }$ (the $%
f$-level position with respect to the chemical potential $\mu $, which is of
order of the Kondo temperature $T_{\text{K}}$) and $V_{\sigma }$ (the
effective hybridization). The equations for these quantities are presented in Appendix.

The corresponding density of states (DOS) reads%
\begin{equation}
N_{\sigma }(E)=\left( 1+\frac{V_{\sigma }^{2}}{(E-W_{\sigma })^{2}}\right)
\rho \left( E-\frac{V_{\sigma }^{2}}{E-W_{\sigma }}\right)
\end{equation}%
(where $\rho (E)$ ($0<E<D$) is the bare DOS) and is shown in Fig.1. The
total capacity of $N(E)$ band is twice larger than that of $\rho (E)$ band
(i.e., unity per each hybridization subband), since it includes
contributions from both conduction electrons and pseudofermions. One can see
that $N(E)$ reproduces and enhances peculiar features of the bare band. For
realistic (considerably smaller) $V$ values in the Kondo systems the
enhancement can be stronger.

\begin{figure}[htbp]
\includegraphics[width=3.3in, angle=0]{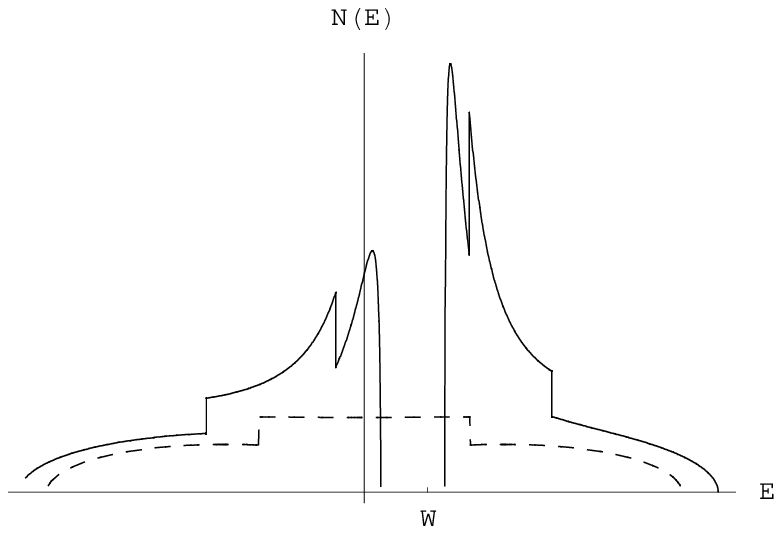}
\caption{The picture of hybridization density of states for  the cubic
lattice in the nearest-neighbor approximation. For clarity, a large value of
$V$ is taken. The dashed line is the bare density of states}
\label{1}
\end{figure}

\section{Half-metallic Kondo ferromagnetism}

On treating ferromagnetic state, we restrict ourselves to the case where the conduction electron
concentration $n<1$ (the results for $n>1$ are obtained by the particle-hole
transformation).

First we discuss the half-metallic ferromagnetic (HMF) solution where the
chemical potential lies in the energy gap for $\sigma =\uparrow $ (Fig.2) so
that
\begin{equation}
W_{\downarrow }>V_{\downarrow }^{2}/(D-\mu ),~-V_{\uparrow }^{2}/\mu
<W_{\uparrow }<V_{\uparrow }^{2}/(D-\mu ),  \label{eq:O.24}
\end{equation}

\begin{figure}[htbp]
\includegraphics[width=3.3in, angle=0]{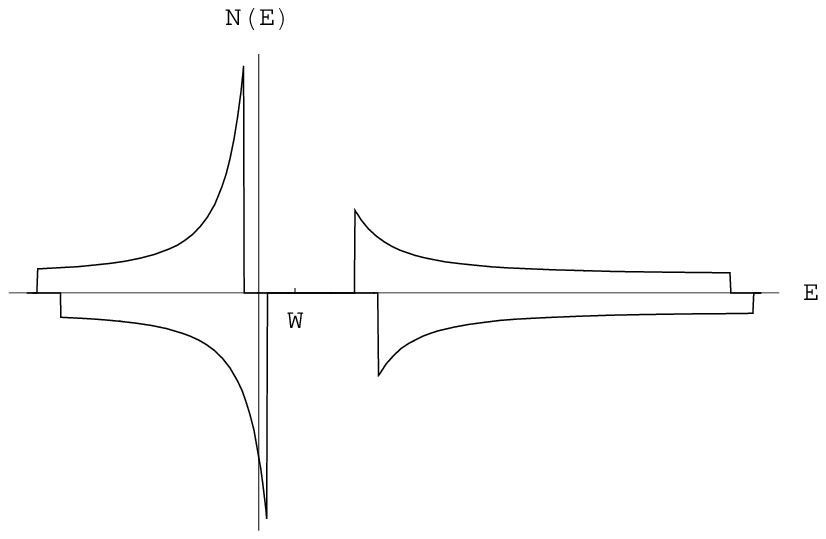}
\caption{The partial spin-up (upper half) and spin-down (lower half) densities of states for the
rectangular bare band ($W_{\downarrow }=W_0$). The vertical line is the
chemical potential for the HFM state. Note that the polarizations of conduction
electrons and pseudofermions are opposite}
\label{2}
\end{figure}

Since the capacity of the hybridization subband equals unity, we have%
\[
n_{\uparrow }^{f}\simeq 1-n/2,~n_{\downarrow }^{f}\simeq n/2,~\bar{S}\simeq
(1-n)/2
\]%
whereas the magnetization of conduction electrons is still small, $%
n_{\uparrow }\simeq n_{\downarrow }\simeq n/2.$ In this state, each
conduction electron compensates one localized spin due to negative sign of 
$s-f$ exchange parameter $I$, and the magnetic ordering is owing to exchange
interaction between non-compensated moments. Such a picture is reminiscent
of the situation in the narrow-band ferromagnet in the Hubbard or $s-d$
exchange model with large intrasite interaction (double-exchange regime). In
our case the bare interaction is small, but effective interaction is large
in the strong coupling regime.

It is worthwhile to mention here an alternative approach based on a Gutzwiller-type variational principle, which was used in Ref.\cite{Fazekas} for low-dimensional Kondo lattices. This approach gives ferromagnetism only at large $|I|$ values (Nagaoka limit) which are beyond the Kondo physics (where a small scale of the Kondo temperature exists).

Using (\ref{vh}), (\ref{wh1}), the second condition in (\ref{eq:O.24}) can be
rewritten as%
\begin{equation}
-M(n)\frac{\mu (2n)-\mu }{\mu }<1-\frac{J_{0}(1-n)}{W_{0}Q(n)}<M(n)\frac{\mu
(2n)-\mu }{D-\mu },  \label{eq:O.26}
\end{equation}%
where $W_{0}$ corresponds to the paramagnetic phase, $J_{0}$ is the
Heisenberg exchange interaction.%
\begin{equation}
M(n)\equiv \left( \frac{V_{\uparrow }}{V_{\downarrow }}\right) ^{2}=\exp
{}\left( \frac{1}{\rho }\int\limits_{\mu (2n)}^{D}\frac{\rho (E)}{E-\mu }%
\,dE\right) .  \label{eq:O.27}
\end{equation}%
\[
Q(n)\equiv \frac{W_{\downarrow }}{W_{0}}=\exp {}\left( -\frac{1}{\rho }%
\int\limits_{\mu (2n)}^{\mu (n+1)}\frac{\rho (E)-\rho }{E-\mu }\,dE\right)
\]

For the rectangular bare band ($\rho (E)=1/D,~0<E<D$) \ we have
\[
M(n)[\mu (2n)-\mu ]/(D-\mu )=1,~Q(n)=1,
\]
and the left-hand inequality in (\ref{eq:O.26}) takes the form
\begin{equation}
~j=J_{0}/W_{0}<\frac{2}{n(1-n)}  \label{eq:O.28}
\end{equation}
It is interesting that, as well as for a Heisenberg ferromagnet, the
solution exists for any small $J_{0}$ and can be stabilized by arbitrarily
small exchange energy (of course, the situation can become different
provided that we take into account fluctuations and the corresponding contributions to the total energy).

The character of solutions can change for an arbitrary form of $\rho (E)$.
To ensure the large energy gap for spin up states, it is required that $\rho
(E)$ is not small in the interval $[\mu _{2n},D]$ and this interval is not
narrow. In particular, the condition (\ref{eq:O.26}) holds if the electron
concentration is small, especially provided that $\rho (E)\ $\ has
square-root behavior near the band bottom ($M(n)$ becomes large owing to the
factor of $1/\rho $). Thus for small $J_{0}$ the HFM solution can disappear
in some concentration region, but become restored with increasing $J_{0}.$

On the other hand, for some bare DOS and $n$ (where $M(n)[\mu (2n)-\mu
]/(D-\mu )>1$) the HFM state can be formally retained for \textit{negative} $%
J_{0}$. Indeed, since the spin up hybridization gap is larger because of
renormalization of $V_{\sigma }$, there can exist HFM solutions with
positive $\bar{S}$ but negative spin splitting, $W_{\uparrow }<W_{\downarrow
}$ (of course, they are not energetically favorable owing to $f-f$ exchange
interaction). The situation is again connected with negative sign of $s-f$
exchange parameter: polarization of conduction electrons is antiparallel to
the moment of $f$-electrons, and occurrence of magnetization results in a
redistribution of electron density.

Note that the picture of half-metallic magnetism is intimately related to
hybridization character of the spectrum, as well as in intermetallic $d$%
-systems \cite{RMP}.

\section{Weak Kondo ferromagnetism}

Now we consider the ferromagnetic solution with small magnetization $\bar{S}%
~<(1-n)/2$\ where the condition $W_{\sigma }>V_{\sigma }^{2}/(D-\mu )$ holds
for both $\sigma $ and the Fermi level lies in the lower hybridization
subband (below the energy gap), as well as in the non-magnetic case.

We have%
\begin{equation}
\frac{J_{0}\bar{S}}{W}=\frac{W_{\downarrow }-W_{\uparrow }}{W_{\uparrow
}+W_{\downarrow }}  \label{WW}
\end{equation}

Taking into account renormalization of hybridization in our model (see
Appendix), we derive the self-consistent equation for magnetization
\begin{eqnarray}
\frac{J_{0}\bar{S}}{W} &=&{}L(\bar{S},n),  \label{Th} \\
~L(\bar{S},n) &=&\tanh \left( \frac{1}{2\rho _{n}}\int\limits_{\mu (n+1-2%
\bar{S})}^{\mu (n+1+2\bar{S})}dE~\frac{\rho (E)-\rho}{E-\mu}%
\,\right) ,  \nonumber
\end{eqnarray}%
($\rho=\rho(\mu)$) and the expression for the renormalized Kondo temperature ($f$-level energy)

\begin{eqnarray}
~W &=&W_{0}P(\bar{S},n),~ \\
P(\bar{S},n) &=&\frac{1}{2}\sum_{\sigma }\exp \left( \frac{1}{\rho }%
\int\limits_{\mu (n+1)}^{\mu (n+1+2\sigma \bar{S})}\frac{\rho (E)-\rho }{%
E-\mu }\,dE\right)   \nonumber
\end{eqnarray}%
The non-saturated solution is smoothly joined with the HMF solution of
previous Section at
\[
n+1-2\bar{S}=2n,~n+1+2\bar{S}=2
\]%
In terms of the quantity $W_{0}$,
\[
W_{0}\simeq V_{0}^{2}/\delta _{n},~\delta _{n}=\mu _{n+1}-\mu _{n},~\mu
_{n}=\mu (n),~
\]%
(which is more convenient for constructing phase diagram) we have the
equation%
\begin{equation}
~j\bar{S}=P(\bar{S},n){}L(\bar{S},n),~j=J_{0}/W_{0}.  \label{eq:O.23}
\end{equation}%
\qquad


Solutions with $\bar{S}\neq 0$ may occur if both the left and right-hand
side of (\ref{Th}) are of order unity, i.e. $J_{0}\sim W_{0}$. However, in
fact the conditions are rather restrictive (note that the situation is
somewhat similar to Hubbard-I approximation where strong dependence of
magnetism criterion on on bare DOS occurs \cite{Hubbard-I:1963}). In
particular, the equation (\ref{Th}) has no non-trivial solutions for $\rho
(E)=\mathrm{const}$: the magnetization can occur only due to energy
dependence of $\rho .$

A necessary (but not sufficient) condition for existence of ferromagnetism
with small $\bar{S}~$is%
\begin{equation}
k_{n}=\left. \frac{dL(\bar{S},n)}{d\bar{S}}\right\vert _{\bar{S}=0}=\frac{1}{%
\rho _{n+1}\rho _{n}}\frac{\rho _{n+1}-\rho _{n}}{\mu _{n+1}-\mu _{n}}>0
\label{K}
\end{equation}%
where $\rho _{n}=\rho (\mu _{n}),~\mu _{n}^{\prime }=1/(2\rho _{n})$.
Therefore, as compared to half-metallic ferromagnetism which is governed by
the global behavior of $\rho (E)$, the small $\bar{S}$ ferromagnetism
criterion is determined by $\rho (E\simeq \mu _{n+1}).$The situation for
existence of small $\bar{S}$ solutions is more satisfactory at larger $n$
(e.g., near half-filling) and in the presence of high narrow peaks (at the
same time, such peaks influence weakly the conditions (\ref{eq:O.26})).

\begin{figure}[htbp]
\includegraphics[width=3.3in, angle=0]{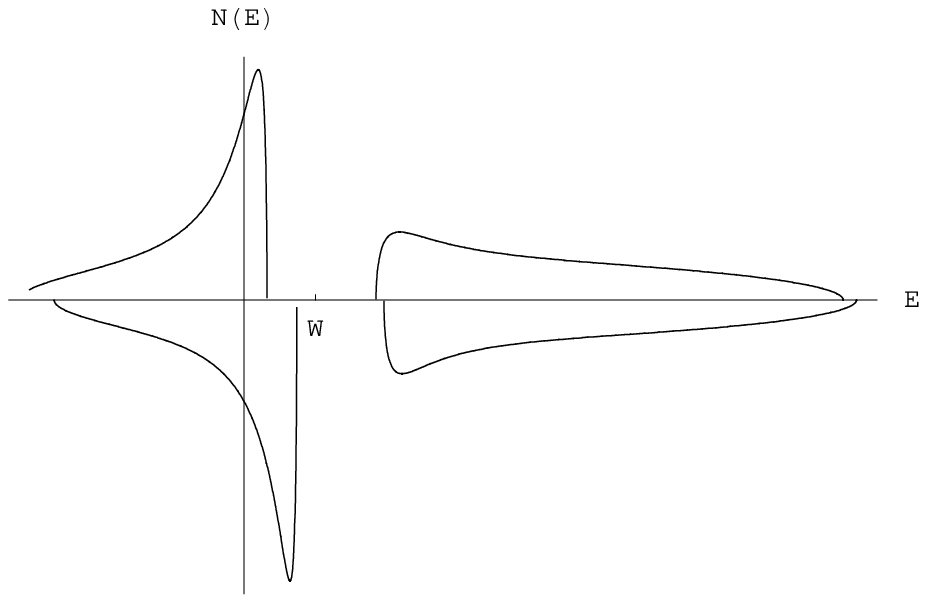}
\caption{The partial spin-up (upper half) and spin-down (lower half)
 densities of states in the case
of semielliptic bare band. The vertical line is the chemical potential for
the non-saturated solution}
\label{3}
\end{figure}

The equation for the ferromagnetic phase at $\bar{S}\rightarrow 0$ is $k=j,$
so that the non-saturated solution starts from $j=k$. Provided that HMF
solution exists for a given $n$, $\bar{S}$ increases with decreasing $j$ up
to the point where $\bar{S}=(1-n)/2$ ($\mu (n+1+2\bar{S})=D$ and the Fermi
level reaches the upper edge of lower hybridization band), $j$ being finite
at this point (the spin splitting 2$J_0\bar{S}$ should remain finite). Thus
this solution exists in a restricted (both from above and below) and even
rather narrow $j$ interval.

The situation changes if HMF solution does not exist at given $n$. Then the
non-saturated solution can exist at arbitrarily small $j$ and even in the
unphysical $j<0$ region where $\bar{S}$ reaches $(1-n)/2$ with increasing $%
|j|$. Such an unusual behavior (in particular, decrease of the moment with
increasing $j$) is connected with a peculiar nature of the non-saturated
solution. In this state we have two competing tendencies.
The growth of exchange splitting $2J_0 \bar{S}$ with $j$ drives the chemical potential into the energy gap, but the sharp pseudofermion peak prevents its own crossing and remains above the Fermi level (Fig.3). Thus $\bar{S}$ should decrease.

In this connection, we can refer to discussion of a sudden jump of the Fermi
surface and of an extra Kondo destruction energy scale for various phases in
Kondo lattices \cite{Si}.

The illustrations based on numerical solution are given in the next Section.
To consider qualitatively the solutions with small moments we \ can perform
the expansion in $\bar{S}$,%
\begin{eqnarray}
~L_{n}(\bar{S}) &=&k_{n}\bar{S}-a_{n}\bar{S}^{3}+...  \label{L} \\
a_{n} &=&[\rho _{n+1}^{\prime \prime }/\delta _{n}-\rho _{n+1}^{\prime
}/\delta _{n}^{2} \\
&&+2(\rho _{n+1}-\rho _{n})/\delta _{n}^{3}]/(12\rho _{n+1})+b_{n}/2
\nonumber
\end{eqnarray}%
where the $\bar{S}$-cubic contributions which come from the expansion of the
chemical potential are similar to those in the usual Stoner theory,%
\begin{eqnarray}
\mu (n+1+2\bar{S})-\mu (n+1-2\bar{S}) \\ \nonumber
=2\bar{S}/\rho _{n+1}+b_{n}\bar{S}%
^{3}+...  \label{mu} \\
b_{n} =(3\rho _{n+1}^{\prime 2}/\rho _{n+1}^{5}-\rho _{n+1}^{\prime \prime
}/\rho _{n+1}^{4})/3  \nonumber
\end{eqnarray}%
Taking into account the renormalization of $W,$%
\begin{eqnarray}
P(\bar{S},n) &=&1-c_{n}\bar{S}^{2}, \\
c_{n} &=&[(\rho _{n+1}-\rho _{n})/\delta _{n}  \nonumber \\
&&+\rho _{n+1}^{\prime }(1-2\rho _{n}/\rho _{n+1})]/(2\delta \rho \rho
_{n+1}^{2}),
\end{eqnarray}%
we can estimate the magnetic moment as%
\begin{equation}
\bar{S}=\sqrt{(k_{n}-j)/(a_{n}+c_{n}k_{n}+k_{n}^{3}/3)}
\end{equation}

Thus it is favorable for the small-moment ferromagnetism that $\rho
_{n+1}^{\prime \prime }<0~$and the derivatives are large in absolute value.
For example, this takes place when there is a narrow peak in bare DOS at (or
somewhat higher) $\mu _{n+1}$, so that magnetic ordering shifts the peak
from the Fermi level. Under these conditions, magnetic ordering can occur
near electron concentrations $n$ corresponding to the peak position $E_{1}$,
i.e. $\mu _{n+1}\simeq E_{1}.$

To demonstrate this, we consider in the next Section also the model bare
density of states where, besides the symmetric smooth part $\rho _{0}(E)$,
the bare DOS has near $\mu _{n+1}$ a narrow Lorentz peak with the width $%
\Gamma ,$%
\begin{equation}
\rho (E)=\rho _{0}(E)+\frac{h\Gamma }{(E-E_{1})^{2}+\Gamma ^{2}}  \label{Lor}
\end{equation}%
One can see from Fig.4 that the Lorentz peak is considerably enhanced and
narrowed by hybridization effects. Such an electronic structure is favorable
for ferromagnetism since a part of the spin up peak is below the Fermi level
which is in a local minimum.

\begin{figure}[htbp]
\includegraphics[width=3.3in, angle=0]{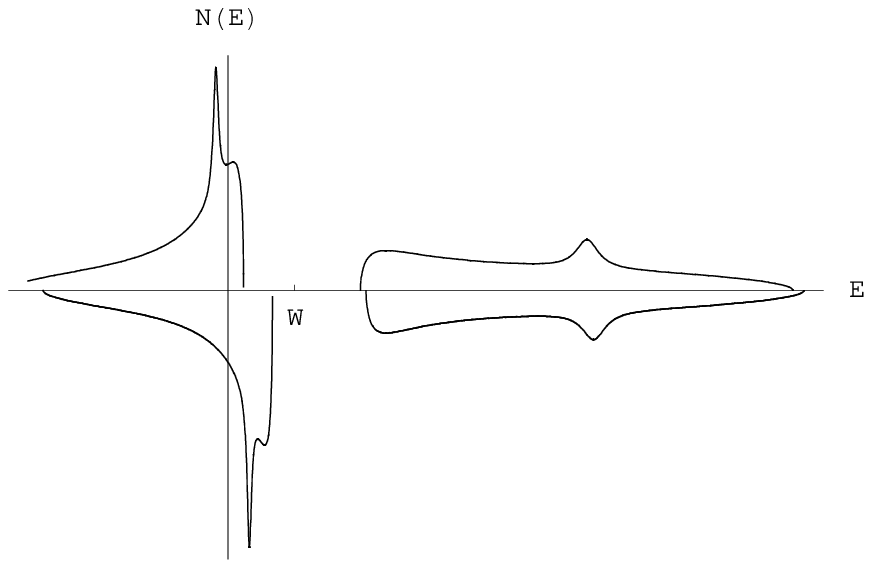}
\caption{ The partial spin-up (upper half) and spin-down (lower half)  densities of states for  the
semielliptic bare band with the Lorentz peak. The vertical line is the
chemical potentia}
\label{4}
\end{figure}


\section{Numerical calculation results}

In the mean-field approximation, the total energy of the magnetic state is always lower than that of
non-magnetic Kondo state,%
\begin{equation}
\mathcal{E}-\mathcal{E}_{\text{non-mag}}=-J_{0}\overline{S}^{2}
\label{eq:O.29}
\end{equation}%
(see Appendix). Thus the formation of the state of a Kondo ferromagnet is
energetically favorable. For the solututions with small moments the energy
gain is smaller than for the half-metallic state, and the latter dominates
provided that the corresponding solution does exist.

The Kondo ferromagnetic state energy should be also compared with the energy
of the usual (Heisenberg) ferromagnetic state with the Kondo effect being
suppressed ($W=0$, $\bar{S}=1/2$). The latter state becomes energetically
favorable at rather large
\begin{equation}
j>j_{c}=1/(1/4-\overline{S}^{2}).
\end{equation}%
At the critical point, a first-order magnetic transition should take place.
Since the corresponding $j$ values are large, we will not show them
on the plots below. We also so not discuss the phase boundaries
corresponding to the left-hand inequality in (\ref{eq:O.26}) (for the cases
considered below, this inequality starts to work and the HFM solution
disappears nearly at the same values, $j>5$).

The universal functions $L(\bar{S},n),~P(\bar{S},n),~M(n)$ and $Q(n)$,
determining the magnetic solutions, depend on a bare density of states only
and can be calculated for concrete lattices.

\begin{figure}[htbp]
\includegraphics[width=3.3in, angle=0]{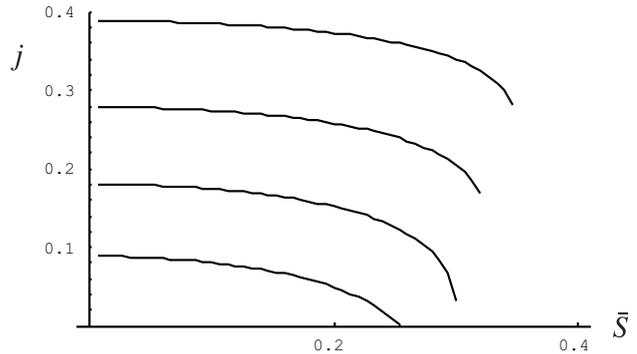}
\caption{The magnetization plots for the semielliptic bare band according to
Eq.(\protect\ref{eq:O.23}) for $n=0.3,0.35,0.4,0.45$ (from above to below) }
\label{5}
\end{figure}

For simple lattices with a symmetric bare DOS, the non-saturated solution
can exist only for $n<1/2$ (where $\rho _{n+1}<\rho _{n}$). In particular,
for the semielliptic bare band with%
\[
\rho _{0}(E)=\frac{4}{\pi D^2}\sqrt{D^{2}-(2E-D)^{2}}
\]%
$\allowbreak $ at $J_{0}\rightarrow 0$ ferromagnetism disappears for $n>0.5$%
, and saturated ferromagnetic solution occurs for $n<0.42$. One can see from
Fig.5 that for a given $n$ the value of $j$ depends very weakly on the
moment and changes in a rather narrow interval. This interval reaches zero
for $0.5>n>0.42$ where only the non-saturated solution exists at $%
j\rightarrow 0$. At the same time, the HFM solution at $n>0.42$ can be
restored by rather small $j$ (Fig.6).

\begin{figure}[htbp]
\includegraphics[width=3.3in, angle=0]{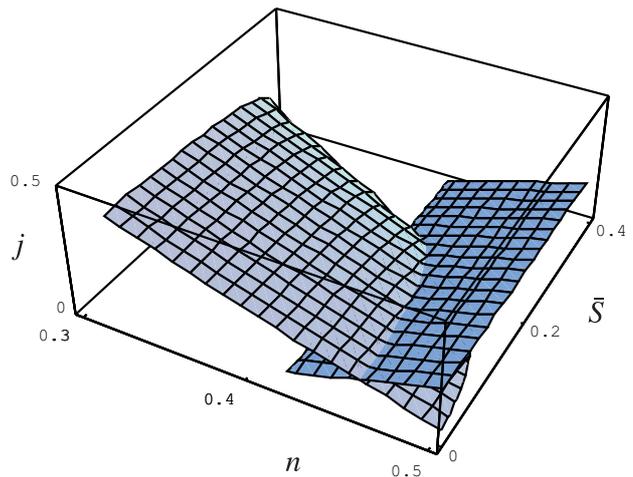}
\caption{(color online) The surface plot of the non-saturated solution for
the semielliptic bare band according to Eq.(\protect\ref{eq:O.23}) (the
left-hand sheet). The HFM solution exists above the right-hand sheet
corresponding to the right-hand inequality Eq.(\protect\ref{eq:O.26})}
\label{6}
\end{figure}

For the square lattice in the nearest-neighbor approximation (where bare
DOS has a logarithmic singularity at the band centre) the HMF solution
exists at $n<0.52.$ The non-saturated solution still exists at $n<0.5,~$i.e.
inside the stability region of the HMF state (note that the singularity with
$\rho _{n+1}^{\prime \prime }>0$ does not favor small moment
ferromagnetism). Thus the HMF state is always more favorable, unlike the
semielliptic band case.

\begin{figure}[htbp]
\includegraphics[width=3.3in, angle=0]{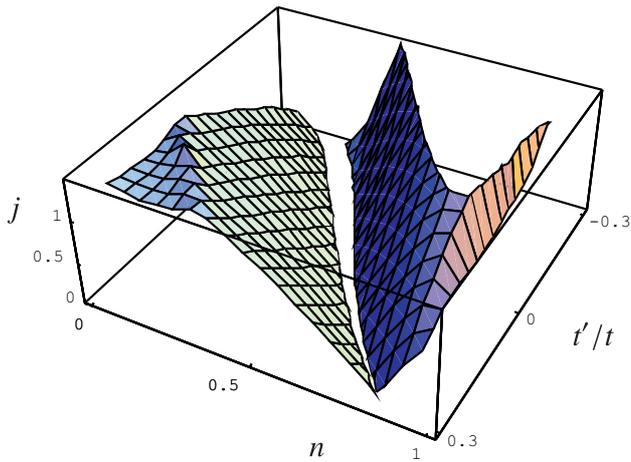}
\caption{(color online) The $\bar{S}\rightarrow 0$ boundary of the
non-saturated solution for the square lattice (the left-hand sheet) and
the boundary of the HFM solution (this exists left of the right-hand sheet)}
\label{7}
\end{figure}

It is instructive to trace the influence of DOS singularity movement on the
phase diagram in the case of square lattice with account of nearest and
next-nearest transfer integrals, $t$ and $t\prime $. Fig.7 demonstrates that
the ferromagnetic region grows for $t^{\prime }>0$ (the singularity is
shifted to the band top) and diminishes for $t^{\prime }<0$ (the singularity
is shifted to the band bottom). However, the region of the non-saturated
solution never goes beyond the region of the the HMF state.

\begin{figure}[htbp]
\includegraphics[width=3.3in, angle=0]{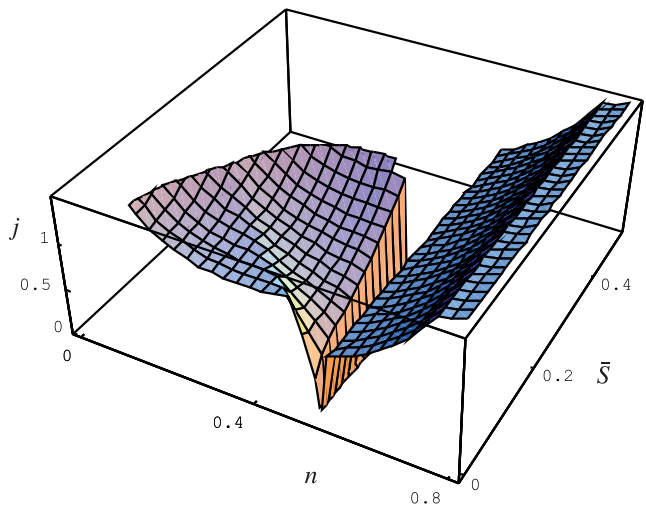}
\caption{(color online) The surface plot of the non-saturated solution for
the simple cubic lattice in the nearest-neighbor approximation (the
left-hand sheet). The HFM solution exists left of the right-hand sheet}
\label{8}
\end{figure}

For the simple cubic lattice (a more weak) Van Hove singularity in bare DOS
(see Fig.1) favours existence of saturated HFM solution, and the
corresponding critical concentration at $j=0$ increases up to 0.5 (Fig.8).
Thus the non-saturated state is again not energetically favorable.

\begin{figure}[htbp]
\includegraphics[width=3.3in, angle=0]{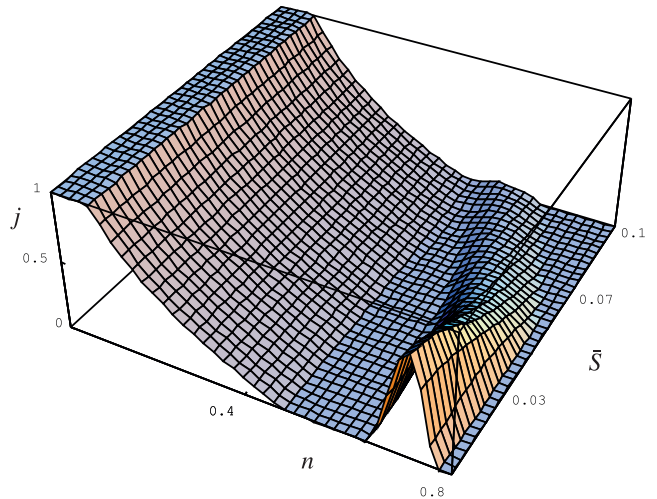}
\caption{(color online) The surface plot of the non-saturated solution for
the semielliptic band with the Lorentz peak (\protect\ref{Lor}) with $\Gamma
=0.01D,~E_{1}=0.8D,~h=0.01$}
\label{9}
\end{figure}

The non-saturated solution for the bare DOS with a peak (\ref{Lor}) is shown
in Fig.9. One can see that small-moment magnetism occurs in a rather wide
region near the $n$ values where $\mu _{n+1}=E_{1}$. Note that the HMF state
does not exist in this concentration region, at least at small $j$. This
case is most interesting from the point of view of analyzing experimental
data on the ferromagnetic Kondo systems (see Introduction).

\section{Conclusions}

We have demonstrated that the pseudofermion ferromagnetism in the Kondo
lattices is a complicated phenomenon which has features of both itinerant
electron and localzed moment magnets. Under certain conditions, magnetic
instability can occur at very small $J$ (even in comparison with $T_{K}$),
which is characteristic for the Heisenberg systems. On the other hand,
magnetic ordering is highly sensitive to electronic structure, as well as
for itinerant systems. We have presented a rather simple description in
terms of the the ratio $j \sim J_{0}/T_{K}$ and universal functions that do not
include the $s-f$ exchange parameter $I$ (or effective hybridization), but
depend only on bare electron DOS $\rho (E)$.

The Kondo ferromagnetism has a number of peculiarities in comparison with
the Stoner picture for usual itinerant systems. In particular, the
dependence of ferromagnetism criterion on the bare density of states turns
out to be different and more complicated. The reason is that $\sigma $%
-dependence of the effective hybridization plays a crucial role for this
criterion.



Owing to hybridization character of spectrum and combined
(electron-pseudofermion) origin of the Fermi surface in the Kondo state,
ferromagnetism is determined mainly by $\rho (\mu _{n+1})$ rather than by $%
\rho (\mu _{n})$.

The hybridization form of the electron spectrum (presence of DOS peaks) in
Kondo lattices is confirmed by numerous experimental investigations: direct
optical data \cite{Mar}, observations of large electron masses in de
Haas-van Alphen effect measurements etc. Sometimes, intermediate valence
(IV) and Kondo regime cannot be clearly distinguished in magnetism formation
since $f$-states play anyway an important role in the electron structure
near the Fermi level. The presence of \textquotedblleft
real\textquotedblright\ hybridization between $s,d$ and $f$-states in IV
systems can induce bare DOS peak which, in turn, will influence
pseudofermion ferromagnetism. A possible example is ferromagnetic CeRh$_{3}$B%
$_{2}$ with high Curie temperature and small saturation moment. However,
neutron scattering analysis for CeRh$_{3}$B$_{2}$ revealed no magnetization
on the rhodium and boron sites, so that ferromagnetism originates from the
ordering of Ce local moments (and not, as has been claimed earlier, from
itinerant magnetism in the Rh 4d band) \cite{Alonso}.

An unsolved problem is investigation of fluctuation effects beyond the
mean-field approximation. In particular, they may destabilize ferromagnetism at small $j$. Simple spin-wave corrections discussed in Ref.\cite{IK91} yield only formally small contributions to the ground state magnetization of order of $j \ln j$ (which are absent in the HMF state). A more consistent consideration of fluctuations can be performed by using the slave-boson approach and the $1/N$ expansion within the periodic Anderson or Coqblin-Schrieffer model \cite{Coleman2}.
An interesting question is the role of fluctuation effects at finite temperatures, which should be even more important than in the ground state (as well as in usual itinerant magnets \cite{26}).


The author is grateful to Prof. M.I. Katsnelson for cooperation and
stimulating discussions.
This work is supported in part by the Programs of fundamental research of the Ural Branch of RAS "Quantum macrophysics and nonlinear dynamics", project No. 12-T-2-1001 and of RAS Presidium "Quantum mesoscopic and disordered structures", project No. 12-P-2-1041.

\section*{Appendix. Mean-field approximation in the pseudofermion
representation}

Making the saddle-point approximation for the path integral describing the
spin-fermion interacting system \cite{Coleman} we reduce the
Hamiltonian of the $s-f$ model to the effective hybridization
model (\ref{eq:O.4}).
After the minimization for this Hamiltonian
we obtain the equations for $W$, the chemical
potential $\mu $ and magnetization $\bar{S}$
\begin{equation}
n_{\sigma }^{f}\equiv \sum_{\mathbf{k}}\langle f_{\mathbf{k}\sigma
}^{\dagger }f_{\mathbf{k}\sigma }\rangle =\frac{1}{2}+\sigma \bar{S};
\label{eq:O.5}
\end{equation}%
\begin{equation}
n=\sum_{\mathbf{k}}\langle c_{\mathbf{k}\sigma }^{\dagger }c_{\mathbf{k}%
\sigma }\rangle .  \label{eq:O.6}
\end{equation}%
\qquad

The quantity $-W$ plays the role of the chemical potential for
pseudofermions, and the numbers of electrons and pseudofermions are
conserved separately. However, there exists an unified Fermi surface, its
volume being determined by the summary number of conduction electrons and
pseudofermions.

After the minimization one obtains the equation for the effective
hybridization $V_{\sigma }$%
\begin{equation}
V_{\sigma }=2I\sum_{\mathbf{k}}\langle f_{\mathbf{k}\sigma }^{\dagger }c_{%
\mathbf{k}\sigma }\rangle .  \label{eq:O.7}
\end{equation}

Diagonalizing the Hamiltonian (\ref{eq:O.4}) by a canonical transformation
\begin{eqnarray}
c_{\mathbf{k}\sigma }^{\dagger } &=&\cos {}(\theta _{\mathbf{k}\sigma
}/2)\alpha _{\mathbf{k}\sigma }^{\dagger }-\sin {}(\theta _{\mathbf{k}\sigma
}/2)\beta _{\mathbf{k}\sigma }^{\dagger },  \nonumber \\
\qquad f_{\mathbf{k}\sigma }^{\dagger } &=&\cos {}(\theta _{\mathbf{k}\sigma
}/2)\beta _{\mathbf{k}\sigma }^{\dagger }+\sin {}(\theta _{\mathbf{k}\sigma
}/2)\alpha _{\mathbf{k}\sigma }^{\dagger },  \label{eq:O.8}
\end{eqnarray}%
with
\begin{eqnarray}
\sin {}\theta _{\mathbf{k}\sigma } &=&\frac{2V_{\sigma }}{E_{\mathbf{k}%
\sigma }},\qquad \cos {}\theta _{\mathbf{k}\sigma }=\frac{t_{\mathbf{k}}-\mu
-W_{\sigma }}{E_{\mathbf{k}\sigma }},  \label{eq:O.9} \\
E_{\mathbf{k}\sigma } &=&[(t_{\mathbf{k}}-\mu -W_{\sigma })^{2}+4V_{\sigma
}^{2}]^{1/2}  \nonumber
\end{eqnarray}%
we obtain the energy spectrum of a hybridization form
\begin{equation}
E_{\mathbf{k}\sigma }^{\alpha ,\beta }=\frac{1}{2}(t_{\mathbf{k}}-\mu
+W_{\sigma }\pm E_{\mathbf{k}\sigma }).  \label{eq:O.10}
\end{equation}%
Then the equations (\ref{eq:O.5})-(\ref{eq:O.7}) take the form
\begin{equation}
n_{\sigma }^{f}=\frac{1}{2}\sum_{\mathbf{k}}\left[ (1-\cos {}\Theta _{%
\mathbf{k}\sigma })n_{\mathbf{k}\sigma }^{\alpha }+(1+\cos {}\Theta _{%
\mathbf{k}\sigma })n_{\mathbf{k}\sigma }^{\beta }\right] ,  \label{eq:O.11}
\end{equation}%
\begin{equation}
n=\frac{1}{2}\sum_{\mathbf{k}\sigma }\left[ (1+\cos {}\Theta _{\mathbf{k}%
\sigma })n_{\mathbf{k}\sigma }^{\alpha }+(1-\cos {}\Theta _{\mathbf{k}\sigma
})n_{\mathbf{k}\sigma }^{\beta }\right] ,  \label{eq:O.12}
\end{equation}%
\begin{equation}
1=-2I\sum_{\mathbf{k}}(n_{\mathbf{k}\sigma }^{\beta }-n_{\mathbf{k}\sigma
}^{\alpha })/E_{\mathbf{k}\sigma }.  \label{eq:O.13}
\end{equation}%
At small $|V_{\sigma }|$, $|W_{\sigma }|$ and $T=0$ we have
\begin{equation}
\cos {}\Theta _{\mathbf{k}\sigma }\simeq \mathrm{sign}{}(t_{\mathbf{k}}-\mu
-W_{\sigma }),  \label{eq:O.14}
\end{equation}%
so that the equations (\ref{eq:O.11}), (\ref{eq:O.12}) are further
simplified.

The edges of the hybridization gaps in spin subbands are given by
\begin{equation}
E_{\mathbf{k}\sigma }^{\alpha }>W_{\sigma }+V_{\sigma }^{2}/\mu ,\qquad E_{%
\mathbf{k}\sigma }^{\beta }<W_{\sigma }-V_{\sigma }^{2}/(W-\mu ).
\label{eq:O.15}
\end{equation}%
Defining the function $\mu (n)$ by
\begin{equation}
n=2\int\limits_{0}^{\mu (n)}\rho (E)\,dE,  \label{eq:O.17}
\end{equation}%
the equation (\ref{eq:O.12}) takes the form $\mu (n)=\mu $, and eqs. (\ref%
{eq:O.11}) and (\ref{eq:O.13}) yield at $W_{\sigma }>V_{\sigma }^{2}/(D-\mu
) $
\begin{equation}
\lambda _{\sigma }\equiv V_{\sigma }^{2}/W_{\sigma }=\mu (n+2n_{\sigma
}^{f})-\mu (n),  \label{eq:O.18}
\end{equation}%
\begin{equation}
1=-2I\int\limits_{0}^{\mu +\lambda _{\sigma }}\frac{\rho (E)}{[(E-\mu
-W_{\sigma })^{2}+4V_{\sigma }^{2}]^{1/2}}\,dE.  \label{eq:O.19}
\end{equation}

In the leading approximation $V_{\sigma }$ does not depend on $\sigma $ and
we have
\begin{equation}
|V_{\sigma }|\sim (DT_{\text{K}})^{1/2},\qquad T_{\text{K}}=D\exp {}\frac{1}{%
2I\rho(\mu) }.
\end{equation}%
\begin{equation}
W\simeq V^{2}/[\mu (n+1)-\mu ]\sim T_{\text{K}}.
\end{equation}%
For $\rho (E)=\mathrm{const}$ the calculations can be performed more
accurately to obtain

\begin{equation}
|V_{\sigma }|^{2}=\mu T_{\text{K}}/2,\qquad W_{0}=\mu \exp {}\frac{1}{2I\rho
}=\frac{nD}{2}\exp {}\frac{1}{2I\rho}.  \label{eq:O.21}
\end{equation}%
To take into account spin polarization one has to calculate the integral in (%
\ref{eq:O.19}) to next-order terms in $1/\ln {}|W/V_{\sigma }|$ \cite{IK91}.
One gets at neglecting $W_{\sigma }$

\begin{equation}
\left( \frac{V_{\sigma }}{V_{0}}\right) ^{2}=\exp \left( \frac{1}{\rho }%
\int\limits_{\mu (n+1)}^{\mu (n+2n_{\sigma }^{f})}\frac{\rho (E)}{E-\mu }%
\,dE\right) .  \label{vv}
\end{equation}%
Using (\ref{eq:O.18}) and (\ref{vv}) we obtain

\begin{equation}
\frac{W_{\sigma }}{W_{0}}=\exp \left( \frac{1}{\rho }\int\limits_{\mu
(n+1)}^{\mu (n+2n_{\sigma }^{f})}\frac{\rho (E)-\rho(\mu)}{E-\mu }\,dE\right)
\label{w}
\end{equation}%
which yields the self-consistent equation for magnetization (\ref{Th}).

In the same way, for the half-metallic state (\ref{eq:O.24}) with $\bar{S}%
=(1-n)/2$ we get%
\begin{eqnarray}
1=-2I\int\limits_{0}^{D}\frac{\rho (E)}{[(E-\mu )^{2}+4V_{\uparrow
}^{2}]^{1/2}}\,dE \nonumber \\ 
=-2I\int\limits_{0}^{\mu (2n)}\frac{\rho (E)}{[(E-\mu )^{2}+4V_{\downarrow
}^{2}]^{1/2}}\,dE,    \label{vh}
\end{eqnarray}
\begin{equation}
\lambda _{\downarrow }=\mu (2n)-\mu (n).  \label{wh1}
\end{equation}

The ground state energy of the Kondo state is given by%
\begin{equation}
\mathcal{E}=\Omega (\mu (n))+J_{0}\overline{S}^{2}+\mu n-W
\end{equation}%
where the second term is introduced to compensate the \textquotedblleft
double-counted\textquotedblright\ terms, $\Omega (\mu (n))=\langle H-\mu
\hat{n}\rangle .~$In our approximation of small $V_{\sigma }^{2}/D$ we
derive
\begin{equation}
\mathcal{E}-\mathcal{E}_{\text{non-mag}}=J_{0}\overline{S}^{2}+\sum_{\sigma
}n_{\sigma }^{f}(W_{\sigma }-W_{0})-W+W_{0}=-J_{0}\overline{S}^{2}
\end{equation}%
(the dependence of effective hybridization $V$ on $\sigma $ does not
influence the magnetic energy since the non-universal hybridization does not
enter $\mathcal{E}$ \cite{IK91}).


\begin{thebibliography}{99}
\bibitem{IK} V.Yu. Irkhin and M.I. Katsnelson, Phys.Rev.B\textbf{56}, 8109
(1997); B\textbf{59}, 9348 (1999).

\bibitem{II} V.Yu. Irkhin and Yu.P Irkhin. Electronic structure, correlation
effects and properties of d- and f-metals and their compounds. Cambridge
International Science Publishing, 2007.

\bibitem{CePt} J. Larrea J., M. B. Fontes, A. D. Alvarenga, E. M.
Baggio-Saitovitch, T. Burghardt, A. Eichler, and M. A. Continentino, Phys.
Rev. B \textbf{72}, 035129 (2005).

\bibitem{CeRuGe} S. Sullow, M.C. Aronson, B. D. Rainford, and P. Haen, Phys. Rev. Lett. {\bf 82}, 2963 (1999).

\bibitem{CeAgSb} V. A. Sidorov, E. D. Bauer, N. A. Frederick, J. R.
Jeffries, S. Nakatsuji, N. O. Moreno, J. D. Thompson, M. B. Maple, and Z.
Fisk, Phys. Rev. B \textbf{67}, 224419 (2003)

\bibitem{CeRu2Ga2B1} R. E. Baumbach, X. Lu, F. Ronning, J. D. Thompson and
E. D. Bauer, J. Phys.: Condens. Matter \textbf{24}, 325601 (2012).

\bibitem{CeRu2Ga2B} H. Sakai, Y. Tokunaga, and S. Kambe, R. E. Baumbach, F.
Ronning, E. D. Bauer, and J. D. Thompson, Phys. Rev. B \textbf{86}, 094402
(2012). 

\bibitem{CeIr2B2} A. Prasad, V. K. Anand, U. B. Paramanik, Z. Hossain, R.
Sarkar, N. Oeschler, M. Baenitz, and C. Geibel, Phys. Rev. \textbf{B 86},
014414 (2012).

\bibitem{CeNiSn} B. Chevalier, M. Pasturel, J.-L. Bobet, R. Decourt, J.
Etourneau, O. Isnard, J. Sanchez Marcos, J. Rodriguez Fernandez, Journal of
Alloys and Compounds \textbf{383}, 4 (2004).

\bibitem{CeRuPO} C. Krellner, N. S. Kini, E. M. Bruning, K. Koch, H. Rosner,
M. Nicklas, M. Baenitz, and C. Geibel, Phys. Rev. B \textbf{76}, 104418
(2007).

\bibitem{CeRuSi2} V. N. Nikiforov, M. Baran, A. Jedrzejczak, V. Yu. Irkhin,
European Phys. Journ. B \textbf{86,} 238 (2013).

\bibitem{Ce4Sb3} V. N. Nikiforov, V. V. Pryadun, A. V. Morozkin, V.Yu. Irkhin,
Physica B \textbf{443}, 80 (2014).

\bibitem{Coleman1} P. Coleman, Heavy Fermions: electrons at the edge of
magnetism, In: Handbook of Magnetism and Advanced Magnetic Materials. Ed. H.
Kronmuller and S. Parkin. Vol 1: Fundamentals and Theory. Wiley, p. 95
(2007).

\bibitem{Coleman} P. Coleman, N. Andrei, J. Phys.: Condens. Matter. \textbf{1%
}, 4057 (1989).

\bibitem{IK91} V. Yu. Irkhin, M. I. Katsnelson, J. Phys.: Cond. Mat.\textbf{2}, 8715 (1990);
Z. Phys. B \textbf{82}, 77 (1991).

\bibitem{Nolting}
C. Santos and W. Nolting, Phys. Rev. B \textbf{65}, 144419 (2002); 
S. Henning and W. Nolting, Phys. Rev. B \textbf{79}, 064411 (2009).

\bibitem{Fazekas} P. Fazekas and E.  Mueller-Hartmann, Z.Phys. B \textbf{85}, 285 (1991).

\bibitem{RMP} M. I. Katsnelson, V. Yu. Irkhin, L. Chioncel, A. I.
Lichtenstein, R. A. de Groot, Rev. Mod. Phys. \textbf{80}, 315 (2008).

\bibitem{Hubbard-I:1963} J. Hubbard, Proc. Roy. Soc. \textbf{A276}, 238
(1963).

\bibitem{Si} Q. Si, J. H. Pixley, E. Nica, S. J. Yamamoto, P. Goswami, R. Yu, S. Kirchner,
arXiv:1312.0764.

\bibitem{Mar} F. Marabelli, P. Wachter, J. Magn. Magn. Mater. \textbf{70},
364 (1987).

\bibitem{Alonso} J. A. Alonso, J.-X. Boucherle, F. Givord, J. Schweizer, B.
Gillon and P. Lejay, J. Magn. Magn. Mater. \textbf{177-181}, 1048 (1998);
F Givord, J-X Boucherle, E Lelievre-Berna and P Lejay, J. Phys.: Cond. Mat.
\textbf{16}, 1211 (2004).

\bibitem{Coleman2}
N. Read,   D.M. Newns, J. Phys. C\textbf{16}, 3273 (1983);
P. Coleman, Phys. Rev. B\textbf{35}, 5072 (1987).

\bibitem{26}
T. Moriya, Spin fluctuations in itinerant electron magnetism, Springer, 1983.

\end{thebibliography}
\end{document}